\documentclass{ws-ijmpcs}

\begin{document}

\markboth{W\l odek Bednarek}
{EXPECTED GAMMA-RAY EMISSION FROM X-RAY BINARIES}

%%%%%%%%%%%%%%%%%%%%% Publisher's Area please ignore %%%%%%%%%%%%%%%
%
\catchline{}{}{}{}{}
%
%%%%%%%%%%%%%%%%%%%%%%%%%%%%%%%%%%%%%%%%%%%%%%%%%%%%%%%%%%%%%%%%%%%%

\title{EXPECTED GAMMA-RAY EMISSION FROM X-RAY BINARIES
%\footnote{For the title, try not to use more than 
%3 lines. Typeset the title in 10 pt roman, uppercase and 
%boldface.}
}

\author{W\l odek Bednarek
%\footnote{
%Typeset names in 8 pt roman, uppercase. Use the footnote 
%to indicate the
%present or permanent address of the author.}
}

\address{Department of Astrophysics, University of \L\'od\'z\\
90-236 \L \'od\'z, ul. Pomorska 149/153, Poland
%\footnote{State completely without abbreviations, the
%affiliation and mailing address, including country. Typeset 
%in 8 pt italic.}\\
bednar@uni.lodz.pl}

\maketitle

\begin{history}
\received{Day .09.2011}
\revised{}
\end{history}

\begin{abstract}
It is at present well known that conditions at some massive binary systems allow acceleration of particles and production  of the GeV-TeV $\gamma$-rays. However, which particles are responsible for this emission  and what radiation processes are engaged is at present not completely clear. We discuss what parameters can determine the acceleration process of particles and high energy radiation produced by them within massive binary systems.   
\end{abstract}

\ccode{PACS numbers: 97.80.-d, 07.85.-m}

\section{Introduction}	

The high energy $\gamma$-ray emission from the binary systems containing energetic pulsars has been considered since late 70-ties
(Cyg X-3~\cite{big77,ve82} or LS I +61 303~\cite{mt81}). However, the first enthusiastic reports on the positive $\gamma$-ray signals in the TeV-PeV energies from a few binaries (e.g. Her X-1, Vela X-1, Cyg X-3, ...) have not been confirmed by more reliable observations with the Whipple Cherenkov telescope~\cite{wee92}.
These likely false reports initiated theoretical investigation of possible propagation of $\gamma$-rays injected into the volume of the binary system~\cite{ps87,mos93}.
The situation changed with the launch of the EGRET telescope
on board of Compton GRO. A few binaries have been found in the large error boxes of the EGRET GeV $\gamma$-ray sources (e.g.
LS 5039~\cite{par00}, Cyg X-3~\cite{mo97},
LS I +61 303~\cite{tho95}, Cen X-3~\cite{ve97}). These claims initiated more detailed studies of the production and propagation of $\gamma$-rays within the binary systems such as, e.g. simulations of the $\gamma$-ray production in the anisotropic Inverse Compton $e^\pm$ pair cascade in the stellar radiation~\cite{bed97,bed00}.

The breakthough came with the discoveries of $\gamma$-ray binaries at TeV energies by the modern Cherenkov telescopes (LS2883/PSR1259~\cite{ah05a}, LS 5039~\cite{ah05b}, LS I +61 303~\cite{al06}.
These discovaries initiated efficient investigation of different scenarios for the $\gamma$-ray production in the massive binary systems.
In this paper I discuss what parameters of the binary systems can influence the acceleration of particles and subsequent production of GeV-TeV $\gamma$-rays in these systems.

\section{Gamma-ray observations of binary systems}

The first binary system containing an energetic pulsar LS2883/PSR1259 was detected at the TeV energies to be related to the crossing the periastron passage~\cite{ah05a}. 
The TeV emission seemed to be related to the moments of the intersection of the pulsar through the equatorial wind of the massive star (Be type) in this system. GeV emission has been also discovered from this system ~\cite{abdo11}, showing an unexpected brightening at about a month after the periastron passage~\cite{tam11}. 
On the other hand, LS 5039 and LS I +61 303 (also now suspected to contain pulsars due to the characteristic radio morphology, Dhawan et al.~\cite{dha06} and Rib\'o et al.~\cite{ribo08} showed clear modulation of the TeV signals with the periods of the binaries with the maximum emission when the compact objects were not far from the inferior conjunction~\cite{al09,ah06a}. 
A clear correlation of the TeV and X-ray emission has been reported in the case of LS I +61 303 and LS 5039, thus supporting their production by this same population of electrons~\cite{an09,tak09}. 
The GeV $\gamma$-ray emission from these two binaries show clear unticorrelation with the TeV $\gamma$-ray light curve~\cite{ab09a,ab09b}. It is known at present that the TeV emission from at least from LS I +61 303 shows long scale variability during which the $\gamma$-ray light curve changes significantly its shape~\cite{acc11,jog11}. The GeV and TeV spectra of LS 5039 and LS I +61 303 show two distinct components with a transition in the energy range between 10-100 GeV~\cite{ab09a,ab09b}.  Recently, two additional massive binary systems, HESS J0632+057 detected in TeV energies~\cite{ah07,acc09} and 1FGL J1018.6-5856 detected in GeV energies~\cite{cor11}, show modulation of the $\gamma$-ray signal with the periods of the associated binary systems. 

Another very compact binary system, Cyg X-3, belonging to the microquasar class, has been also detected in GeV $\gamma$-rays~\cite{tav09,ab09c}. The $\gamma$-ray emission observed from this system is correlated with the major radio flares~\cite{tav09}. 
The emission shows a modulation with the period of the binary system, with the maximum close to the location of the compact object behind the companion star.  
 Up to now, TeV $\gamma$-ray emission has not been discovered from Cyg X-3, in spite of extensive observations with the MAGIC telescope~\cite{ale10}. In the case of another microquasar, the massive binary system Cyg X-1, there are evidences of transient GeV-TeV $\gamma$-ray emission~\cite{al07,sab10}. The flare of TeV $\gamma$-ray emission, almost instantaneous with a flare in the hard X-rays~\cite{al07,mal08}, has been reported from Cyg X-1. It lasted for less than an hour, around the superior conjunction of the compact object.  Also a separate flare of $\gamma$-ray emission above $100$ MeV, on a time scale of 1-2 days, has been reported recently by the Agile Team~\cite{sab10}. These GeV $\gamma$-ray flares seem to be rather exceptional during the 300 day observation period with the Agile telescope~\cite{dm10}. The $\gamma$-ray emission from other microquasars within binary systems has not been discovered up to now in spite of extensive observations, see e.g. GRS 1915+105~\cite{ac09,sai09}, SS433~\cite{ah05c,hay09,sai09} or Sco X-1~\cite{ale11}.

Only one massive binary system containing two massive stars, Eta Carinae, has been likely detected by the Agile and Fermi-LAT telescopes at energies below $\sim$100 GeV~\cite{tav09b,abdo10,wfl10}.
The high energy emission of Eta Carinae is characterised by an intriguing two-component spectrum which can be approximated by two power laws, first one with an exponential cut-off at a few GeV and the second extending up to $\sim$100 GeV. It has been reported that the second $\gamma$-ray component 
varies with the phase of the binary system, thus supporting the evidence that it has to come from within the binary system~\cite{wal11}. It will be very interesting to search for the possible TeV emission from this binary system.

\section{Conditions within massive binary systems}

\begin{figure}[t]
\vskip 5.truecm
\includegraphics{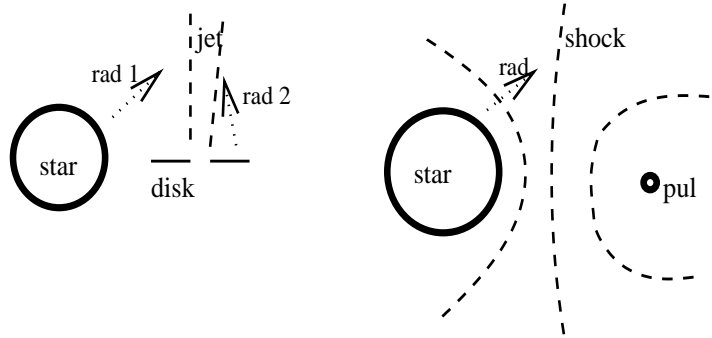}
\caption{The geometry of the interaction of relativistic particles with the radiation field within the binary system. Particles are accelerated at shocks within the jet or at the shock wave which results from the pulsar wind-stellar wind collision. The geometry of the acceleration and the interaction process may/or may not be similar, depending on the parameters of the binary system.}
\label{fig1}
\end{figure}

$\gamma$-ray emission has been detected from two types of massive binary systems. In the first one it is expected that the matter accreting onto a black hole creates an accretion disk. From the inner part of the disk, a jet is launched similarly to the process observed in active galactic nuclei (the so-called microquasar scenario). Particles are expected to be accelerated in the jet, which is usually considered to propagate perpendicularly to the disk plane and to the plane of the binary system (see Fig.~1). In the second case, the binary system contains either two massive stars or an energetic pulsar and a massive star. Both stars produce strong stellar or pulsar winds. 
Particles are expected to be accelerated at the shock structure which appears as a result of the collisions of the winds. They can be also injected from the inner magnetosphere of the pulsar. The geometry of injection of relativistic particles into the binary system in these two scenarios may (or may) not differ significantly. If the pressure of both winds in the binary system is comparable, then the shock structure roughly has a plane shape. However, in the case of the winds with different strengths, the shock can likely bend around one of the stars producing different injection geometry for relativistic particles.

The radiation process by accelerated particles within the binary system can also occur in a complicated environment. After all, the magnetic field around the massive stars can have a complicated structure (see Fig.~2). Depending on the strength of the wind and the stellar rotation velocity, the magnetic field structure is well described by the dipole approximation
($B(r)\propto r^{-3}$). This usually happens only in the region very close to the stellar surface (up to 2 stellar radii in the extreme case). At farther distances, the magnetic field becomes radial ($B(r)\propto r^{-2}$), and finally far away from the star, it is expected a toroidal structure ($B(r)\propto r^{-1}$). The transition between these two last regions is typically at the distance of $\sim$ 10 stellar radii. It is determined by the rotational velocity of the star~\cite{um92}.  

Also the winds produced by the massive stars are not uniform.
For example, the Be type stars (which are expected to be present in LS2883 and LS I +61 303) have dense and slow equatorial winds and fast and rare poloidal winds (see Fig.~2). Therefore, it is expected that the shock structure within the binary system and the conditions for acceleration of particles can change (sometimes even very abruptly) with the orbital phase of the binary system.

\begin{figure}[t]
\vskip 4.6truecm
\includegraphics{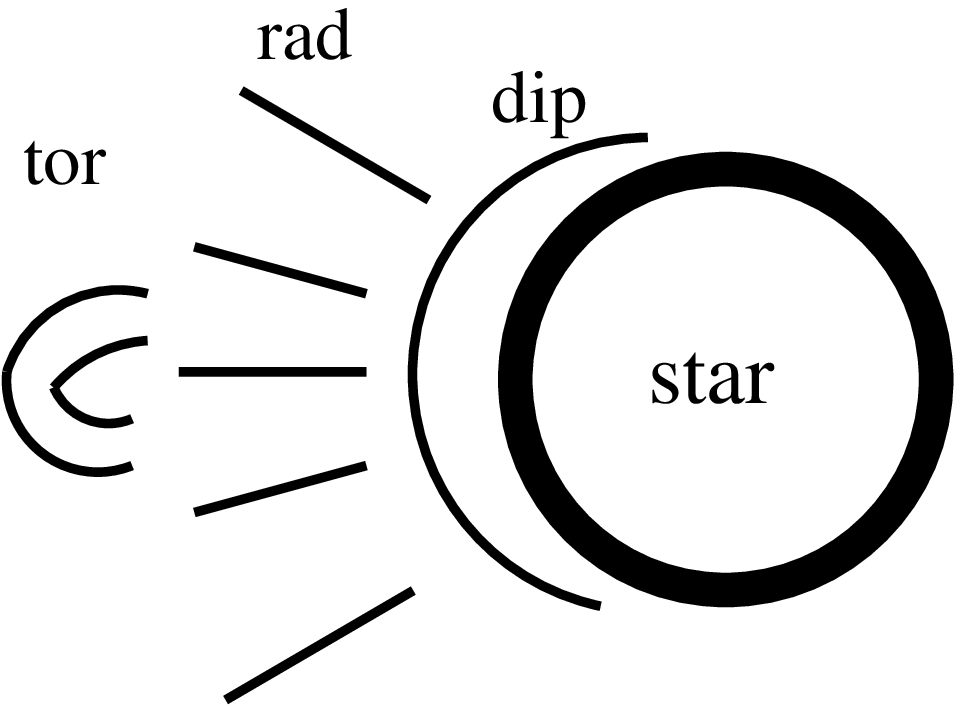}
\includegraphics{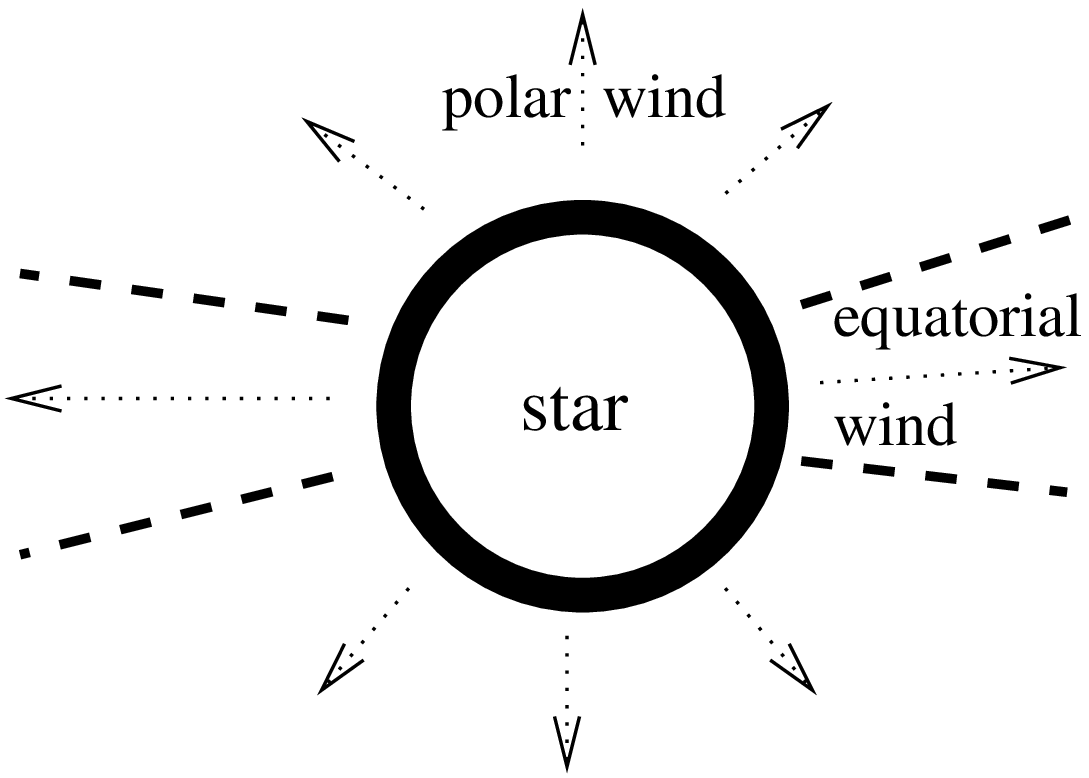}
\caption{The magnetic field around massive stars have a very complicated structure (left figure), close to the star it is dipolar, at larger distances, it is radial, and far away from the star, it becomes toroidal. The winds of Be type stars are strongly inhomogeneous, being slow and dense in the equatorial plane of the rotating stars and fast and rare in the poloidal region (right figure).}
\label{fig2}
\end{figure}

Massive stars in $\gamma$-ray compact binary systems produce a very strong radiation field. $\gamma$-rays with sufficiently large energies are efficiently absorbed on this radiation field.
In fact, detailed calculations of the optical depths for $\gamma$-rays are clearly above unity for the parameters typical for the massive stars~\cite{bed00}. These optical depths strongly depend on the location of the injection place within the binary system and on the propagation angles in respect to the direction towards the massive star (note that for very compact binary systems the dimensions of the stars have to be taken into account). Due to the similarities of the cross sections for the $\gamma-\gamma\rightarrow e^\pm$ absorption process and the Inverse Compton (IC) scattering of soft photons by relativistic electrons, it is expected that the escaping $\gamma$-ray spectra from the binary systems are formed in IC $e^\pm$ pair cascades. These cascades develop in the anisotropic radiation of the massive star since
the injection place of primary particles lays outside the star. As a result of these cascading processes, a dip in the $\gamma$-ray spectrum is expected at energies determined by the temperature of the soft radiation field~\cite{ah06b}.

\section{General scenarios for the IC $e^\pm$ pair cascades}

Since the $\gamma$-ray signals observed from the massive binaries are modulated with their rotational periods, it is expected that $\gamma$-rays are mainly produced within the volume of the binary system where the medium is very inhomogeneous. Below we consider different effects which can influence the emerging $\gamma$-ray spectra.

\subsection{IC $e^\pm$ pair cascades: linear, isotropized, driven by magnetic field}

As we mentioned above, the cascading processes very likely play an
important role in the process of $\gamma$-ray production. However, they can occur in different scenarios.
In the simplest case, the cascade develops through the IC and 
$\gamma-\gamma$ absorption processes along the direction of the first generation of produced $\gamma$-rays (see Fig.~3 on the left). We call this cascade as a linear cascade. Such approximation of the geometry of the cascade is valid provided that the Larmor radii of the secondary $e^\pm$ pairs, produced in the cascade, are much larger than the characteristic IC scattering length (the mean free path for the IC scattering). 
This condition is usually not met within the considered binary systems due to relatively strong magnetic field.
For detailed calculations of the IC $e^\pm$ pair cascade process in this scenario see Cerutti et al.~\cite{cer09}.

In a more realistic IC $e^\pm$ pair cascade process, it is assumed that the secondary cascade $e^\pm$ pairs are completely  isotropized in the place of their creation within the binary system. 
Again, the important criterion for the isotropization condition is the comparison of the Larmor radii of leptons with their IC scattering lengths (both dependent on particle energy and the magnetic field strength and the distance from the star). In this cascade scenario, the next generation of $\gamma$-rays is produced by $e^\pm$ pairs being isotropized in the place of their origin. These leptons produce $\gamma$-rays preferentially in the direction towards the source of the soft radiation field, i.e. the massive star (see Fig.~3 in the center). Some of them can also emerge from the binary system on the opposite side of the star with respect to the location of the source of primary $\gamma$-rays (or electrons). Such type of the cascade has been at first investigated by Bednarek~\cite{bed97,bed00}.

\begin{figure}[t]
\vskip 5.5truecm
\includegraphics{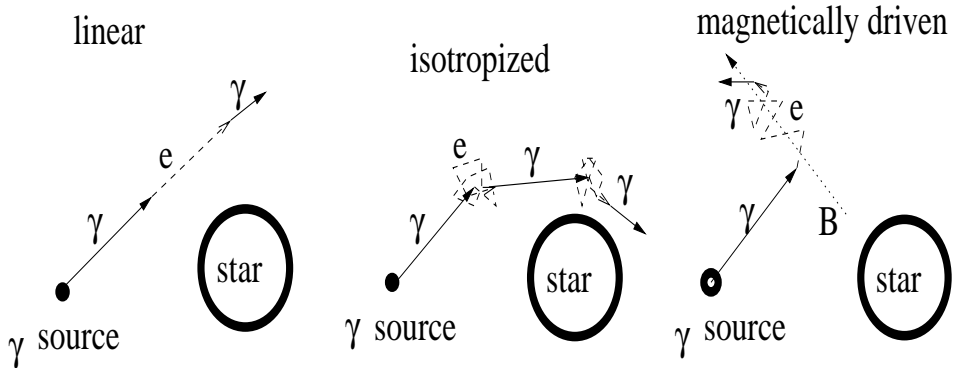}
\caption{Inverse Compton $e^\pm$ pair cascades in the anisotropic radiation field of the massive star can occur
in different geometrical scenarios. Left figure: The linear cascade in which secondary leptons and $\gamma$-rays move along a straight line (magnetic field is too low  to disturb the directions of leptons); Middle figure: Secondary cascade leptons are locally completely isoptopized by the random component of the magnetic field in the stellar wind. Secondary $\gamma$-rays are produced preferentially in the direction towards the massive star. Right figure: Secondary leptons follow the direction of the local magnetic field lines (the most general case) producing the next generation of $\gamma$-rays in a completely different direction than the direction of their parent $\gamma$-rays. }
\label{fig3}
\end{figure}

In the third type of cascade (the most realistic one), secondary cascade $e^\pm$ pairs follow the direction of the local magnetic field lines at the place of their origin (see Fig.~3 on the right). In such a case, the secondary $\gamma$-rays can be produced in completely different directions 
than the direction of propagation of previous generation of $\gamma$-rays. As a result, $\gamma$-rays, escaping from the binary system, can produce a characteristic pattern on the sky around the binary system which depend on the structure of the magnetic field around the star and the geometry of injection of primary electrons (i.e. the angle towards the center of the star, see Fig.~4). Such type of cascade has been at first studied by Sierpowska \& Bednarek~\cite{sb05}.

\begin{figure}[t]
\vskip 10.truecm
\includegraphics{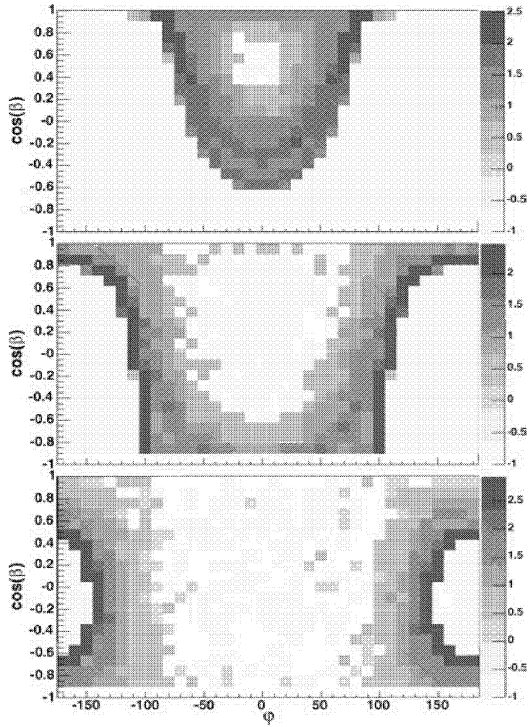}
\includegraphics{Bednarek_1_f4b.eps}
\caption{The distribution of $\gamma$-rays on the sky (darker regions mean more $\gamma$-rays) produced in the IC $e^\pm$ pair cascade in which the secondary leptons follow the local direction of the magnetic field lines (in the place of their origin), see for details Sierpowska \& Bednarek~(2005). The injection angles of primary $\gamma$-rays are equal to: $90^o$ (upper panel), $120^o$ (middle panel), and $150^o$ (bottom panel).}
\label{fig4}
\end{figure}

\subsection{Effects of $e^\pm$ pair energy losses in the magnetic field}

The energy losses of primary electrons and secondary cascade $e^\pm$ pairs on the synchrotron process can play important role 
on the development of the IC $e^\pm$ cascade (see~\cite{bed97,bg07,bos08,kha08}). In fact, by comparing the energy loss rates of electrons on the synchrotron and the IC processes, we can estimate the limit on the magnetic field strength below which the IC losses starts to dominate.
When the ICS occurs in the Thomson regime, the limit on the magnetic field is simply given by the condition $B < 40T_4^2$ G, where $T = 10^4T_4$ K is the surface temperature of the companion star~\cite{bed97}. 
For typical surface temperatures of the stars in these binaries, of the order of $3\times 10^4$ K, this critical value of the magnetic field is equal to $\sim 400$ G, i.e. not so restrictive in the main volume of the binary system. In the Klein-Nishina regime, the condition on the magnetic field strength becomes more restrictive~\cite{bed97}.
Detailed calculations have shown that when the magnetic field in the cascade region is above $\sim 1$ G, then significant energy is transferred from leptons to synchrotron radiation~\cite{bos08,kha08}. As a result,
IC $e^\pm$ pair cascade can start to be inefficient especially at the TeV energies when the IC cross section in the Klein-Nishina regime drops significantly. 
Note that the energy density of the magnetic field around the star drops in the interesting range of distances from its surface as $U_{\rm B}\propto R^{-4}$, whereas the energy density of the stellar radiation drops according to $U_{\rm B}\propto R^{-2}$.
Therefore, the region in which the synchrotron losses of electrons dominate over their IC losses should extend rather relatively close to the companion star. At the distance of $4-15$ stellar radii from the companion star, the magnetic fields are likely to be of the order of $\sim$1 G, in accordance with the calculations~\cite{bed97,bos08,kha08}.

\subsection{Dependence on the shock localization (variable stellar wind)}

The $\gamma$-ray spectra emerging from the anisotropic IC $e^\pm$ pair cascade strongly depend on the shape of the compact object orbit. Therefore, the $\gamma$-ray spectra arriving to the observer should strongly depend on the injection place of electrons
within the binary system which is in fact determined by the structure of the shock and also the orbital shape of the compact object. In Fig.~5, we show how these angles can change in the case of the shocks which bound around the massive star and the compact object. The shape of the shock depends on the pressure of the stellar wind which might change with time and the phase of the compact object (e.g. due to the effects of irradiation of the stellar surface or the anisotropic stellar winds). Therefore, we expect the the $\gamma$-ray light curves from specific binaries (especially those with variable winds, e.g. produced by Be stars) may change from one to another binary cycle. In fact, such time dependent $\gamma$-ray light curves have been observed in the case of LS I +61 303 which likely contain a Be type star.

\begin{figure}[t]
\vskip 6.truecm
\includegraphics{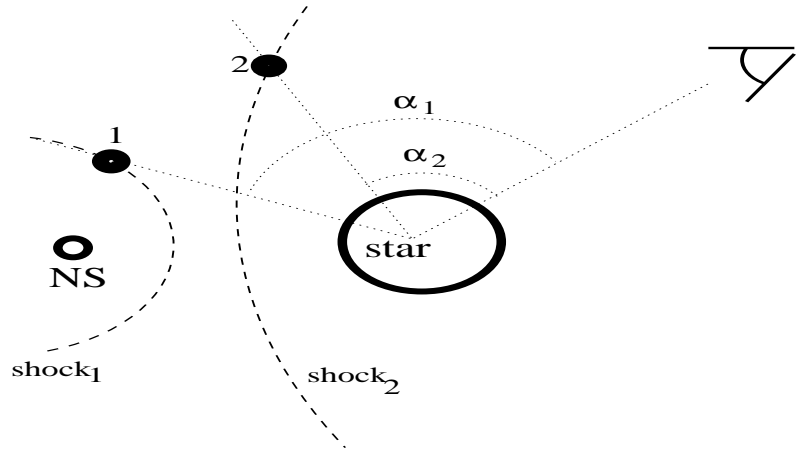}
\caption{The angle between the direction to the observer and the site (a part of the shock region) where electrons are accelerated can change significantly (from the angle $\alpha_1$ to $\alpha_2$), depending on the localization of the shock within the binary system. The location of the shock can shift as a result of the change of e.g. the stellar wind pressure.}
\label{fig5}
\end{figure}

\subsection{Effects of anisotropic stellar/pulsar winds} 

As we  mentioned above, winds of massive stars are expected to be strongly anisotropic. However also the wind produced by the pulsar is expected to be anisotropic~\cite{bk02,vol08}. Therefore, very complicated interaction patterns (shock structures) between such anisotropic winds are expected (for some examples see Fig.~6). 
As a result, depending on relative geometrical orientations, the shock created in the wind collisions can sometimes appear very close to the massive star or to the neutron star~\cite{sb08,st09}.
It is expected that such drastic changes may occur at quite nearby phases of the compact star on its orbit around the massive star.
Thus, it is not very surprising that the $\gamma$-ray light curves from the binary systems can show very irregular features and high level of emission at unexpected phases. In specific phases the shock structure can sometimes appear very close to the star where the conditions for $\gamma$-ray production are the most favourable.

\begin{figure}[t]
\vskip 6.truecm
\includegraphics{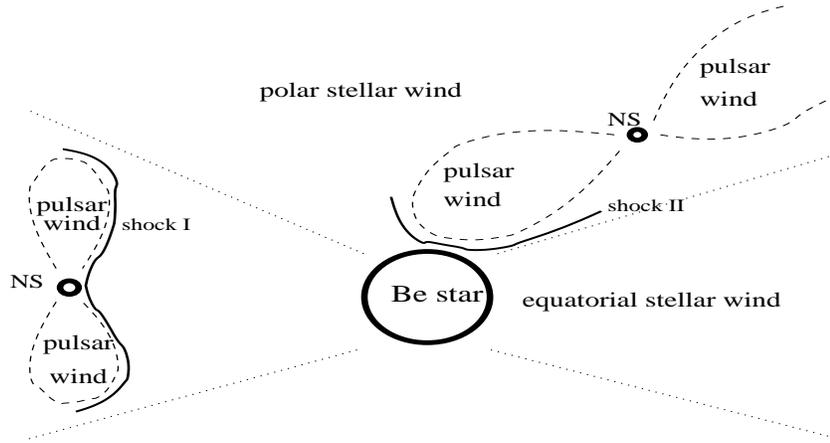}
\caption{Possible geometries of the shock structures (shock I and shock II) which might appear within the binary system as a result of the interaction of axisymmetric pulsar wind with axisymmetric stellar wind which is composed of the equatorial dense and slow wind and poloidal fast and rare wind.}
\label{fig6}
\end{figure}

\subsection{Effects of clumpy stellar wind} 

The winds produced by massive stars are expected to contain dense clumps immersed in a diluted medium filled with matter with different properties (magnetic field strength, density). The accelerated particles can diffuse into such two component medium suffering different propagation and radiation effects. As a result, the produced $\gamma$-ray spectrum may have a complicated structure due to the main energy losses on different radiation processes. Such model has been considered as a possible explanation of the emission features observed in LS I +61 303~\cite{zdzir10}. In the case of microquasars, the interaction of such a dense clumps with the particles accelerated in the jet may also lead to the production of $\gamma$-rays through hadronic processes~\cite{ow09,ar09}.

\subsection{Effects of relativistic boosting of radiation}

The relativistic pulsar wind, after slowing down in the downstream region of the shock, is still expected to move with substantial velocity, of the order of $0.3-0.5c$. This is still enough to collimate at some level the $\gamma$-ray emission produced by relativistic particles in the wind~\cite{dub10}.
In specific conditions, the relativistic flow after the shock is even expected to accelerate to very large Lorentz factors, of the order of $\sim 100$~\cite{b08}. In such a case,
strong spikes of $\gamma$-ray emission should be observed in some phases of the orbit.
Similar relativistic boosting effects, to those observed in active galaxies, has been also considered with the application to microquasars by e.g., Romero et al.~\cite{rom02} and Dubus et al.~\cite{dub10b}.
Such an effects might be responsible for the modulation of the $\gamma$-ray signal with the period of the binary system.

\subsection{Double shock structure - acceleration of two populations of electrons}  

In fact, the interaction of the winds from two stars turn to the production of a double shock structure separated by a contact discontinuity. As a result, the conditions for acceleration of particles on both shocks can differ significantly due to different fluids (characterised by different magnetic field strength, plasma velocities, etc ...).  Therefore, it is expected that particles
are accelerated to different maximum energies on both shock.
As a result, two populations of electrons/hadrons are injected into the binary system. Such a model can provide a nice explanation for the complicated (two component) $\gamma$-ray spectra observed from the binary system of two massive stars, Eta Carinae (see for details Bednarek \& Pabich~\cite{bp11}). In the case of the binary systems containing pulsars, the shock at relativistic wind from the pulsar can accelerate electrons to maximum energies of the order of several TeV~\cite{bed11b}.
On the other hand, particles accelerated at the shock in the stellar wind can reach maximum energies of the order of a few tens GeV. Such a model provides natural explanation of the two component $\gamma$-ray spectra  observed from LS 5039 and LS I +61 303 as a consequence of production of $\gamma$-rays by two populations of electrons with different maximum energies. Note that the kinetic powers of the winds from the massive stars in LS 5039 and LS I +61 303 seem to be comparable to the GeV $\gamma$-ray luminosity. However, the $\gamma$-rays produced in the IC $e^\pm$ pair cascades are highly anisotropic which result in enhanced 
$\gamma$-ray fluxes escaping in specific directions.

\section{Acceleration of electrons and/or hadrons ?}

It has been argued that leptons may have problems with reaching the TeV energies within the binary systems due to the huge energy losses~\cite{ah05b}. In such a case, hadrons can be responsible for GeV-TeV $\gamma$-rays observed from binary systems. Such a hadronic models have been considered in a few papers~\cite{rom03,rom05,kaw04,cher06,th07,ar09,ow09,bp11}.  
In fact, complicated $\gamma$-ray spectra observed from binary systems might be in agreement with the expectations of such models. However, the problem appears whether sufficient amount of energy can be transferred to radiation from hadrons due to their relatively inefficient cooling.
The definitive confirmation of the importance of hadronic radiation processes should come from the observations of the neutrino signals from the $\gamma$-ray binaries.

\section{Conclusion}

Massive binary systems are one of the best defined sites for high energy
processes and $\gamma$-ray production due to the well known geometry, soft radiation field, density of matter and structure of the magnetic field of the massive companion star. However, 
they are still quite complicated objects with complicated geometry. Due to the present lack of precise information, it is difficult to fix the basic conditions in these systems.
Therefore, reliable predictions of the $\gamma$-ray emission features from $\gamma$-ray binary systems are difficult since many effects have to be taken into account such as: geometry of the particle interaction and the binary system in respect to the observer, non-steady medium (aspherical, variable, inhomogeneous winds), different radiation processes, different populations of accelerated particles. It is not surprising that $\gamma$-ray emission features from massive binaries show such variety of 
fine structures and unexpected behaviour.

\section*{Acknowledgments}
This work has been supported by the Polish NCBiR grant: ERA-NET-ASPERA/01/10. 

%\begin{thebibliography}{000} %for 3 digits
%\begin{thebibliography}{00}  %for 2 digits

\end{document}